\documentclass[conference]{IEEEtran}
\IEEEoverridecommandlockouts
\usepackage{cite}
\usepackage{amsmath,amssymb,amsfonts}
\usepackage{algorithmic}
\usepackage{graphicx}
\usepackage{textcomp}
\usepackage{xcolor}
\usepackage{dblfloatfix}
\def\BibTeX{{\rm B\kern-.05em{\sc i\kern-.025em b}\kern-.08em
    T\kern-.1667em\lower.7ex\hbox{E}\kern-.125emX}}

\usepackage{booktabs} 
\usepackage{amsmath}
\usepackage{overpic}

\newcommand*{\everymodeprime}{\ensuremath{\prime}}

\newcommand{\VECTOR}[1]{\ensuremath{\mathbf{#1}}} 

\DeclareMathOperator{\E}{\mathbb{E}}
\DeclareMathOperator{\R}{\mathbb{R}}

\newsavebox\CBox
\def\TEXTBF#1{\sbox\CBox{#1}\resizebox{\wd\CBox}{\ht\CBox}{\textbf{#1}}}

\begin{document} 

\title{CARL: Aggregated Search with Context-Aware Module Embedding Learning}

\author{
    \IEEEauthorblockN{Xinting Huang$^{1}$, Jianzhong Qi$^{2}$, Yu Sun$^{3}$, Rui Zhang$^{2*}$\thanks{$*\,$  Corresponding author}, Hai-Tao Zheng$^{4}$}
    \IEEEauthorblockA{
        \textit{$^{1, 2}$The University of Melbourne}, 
                \textit{$^{3}$Twitter Inc.}, \textit{$^{4}$Tsinghua University}.\\
        $^{1}$xintingh@student.unimelb.edu.au,
        $^{2}$\{jianzhong.qi, rui.zhang\}@unimelb.edu.au,\\
        $^{3}$ysun@twitter.com,
        $^{4}$zheng.haitao@sz.tsinghua.edu.cn
    }
}
\maketitle

\begin{abstract}
Aggregated search aims to construct search result pages (SERPs) from blue-links and heterogeneous modules (such as news, images, and videos). 
Existing studies have largely ignored the correlations between blue-links and heterogeneous modules when selecting the heterogeneous modules to be presented. 
We observe that the top ranked blue-links, which we refer to as the \emph{context}, can provide important information about query intent and helps identify the relevant heterogeneous modules.
For example, informative terms like ``streamed'' and ``recorded'' in the context imply that a video module may better satisfy the query.
To model and utilize the context information for aggregated search, we propose a model with context attention and representation learning (CARL).
Our model applies a recurrent neural network with attention mechanism to encode the context, and incorporates the encoded context information into module embeddings.
The \textit{context-aware} module embeddings together with the ranking policy are jointly optimized under the Markov decision process (MDP) formulation.
%
To achieve a more effective joint learning, we further propose an optimization function with self-supervision loss to provide auxiliary supervision signals.
Experimental results based on two public datasets demonstrate the superiority of CARL over multiple baseline approaches, and confirm the effectiveness of the proposed optimization function in boosting the joint learning process.

\end{abstract}


\section{INTRODUCTION}

In recent years, search engines have become more diverse and specialized.
A wide range of information-seeking tasks can be addressed by specialized search engines, including search for images (Flickr), videos (Youtube), news (BBC), scholarly articles (arXiv.org), etc. 
These specialized search engines are referred to as vertical search systems or simply \textit{verticals}.
Based on a collection of verticals, modern search engines (e.g., Google, Bing) construct \emph{search result pages} (SERP) by aggregated search to fulfill users' information needs across multiple domains. 
For example, an aggregated SERP for query ``reinforcement learning" is shown in Figure 1, in which the vertical results (including scholarly articles, news and videos) are presented as document blocks, i.e., \emph{modules}, alongside the blue-links. 
Regardless of whether the query is issued by a researcher looking for scholarly articles or a student looking for tutorial videos, the desired information is served in this aggregated SERP.

\begin{figure}[!t]
\centering

\includegraphics[width=3.5in]{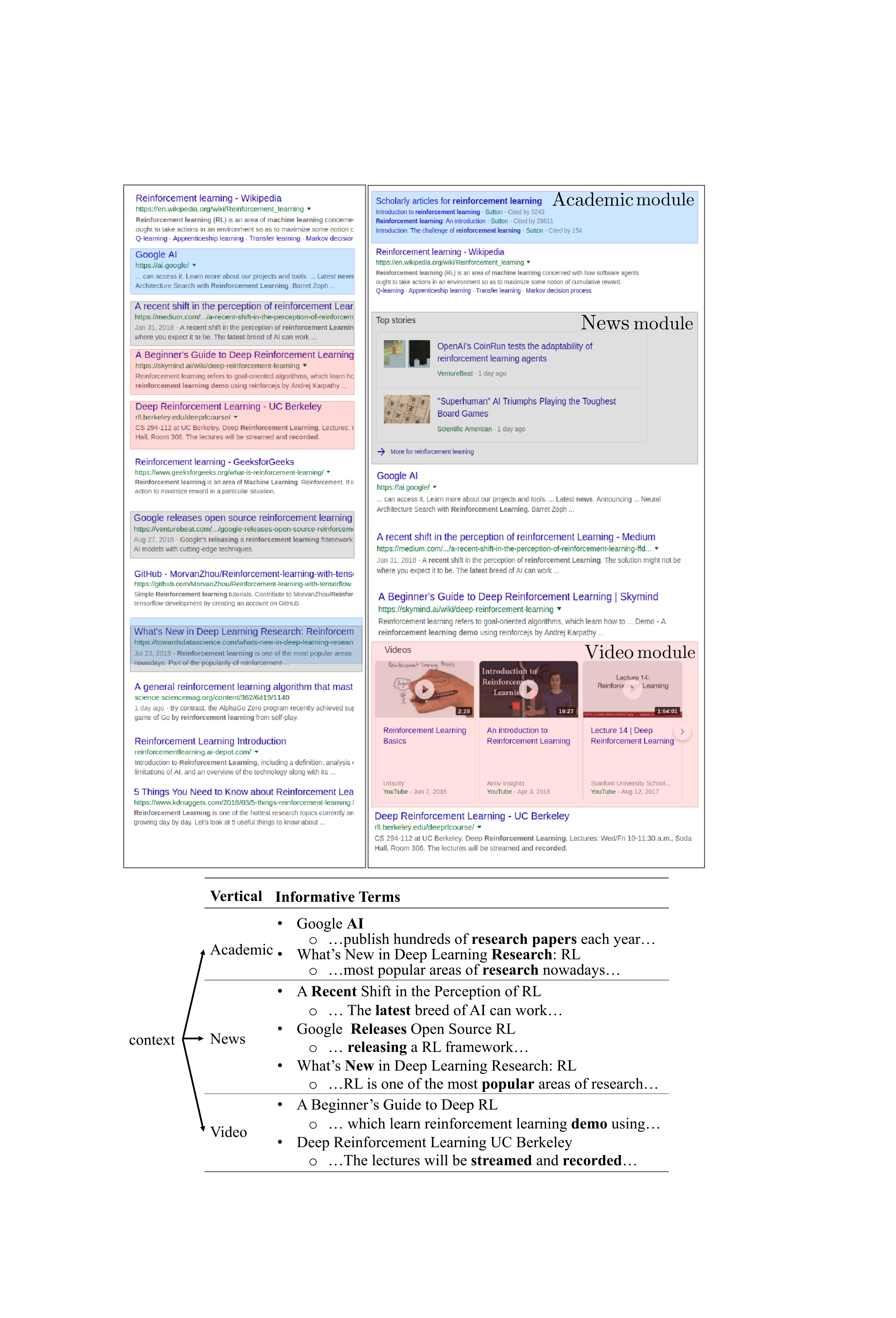}
\caption{ Contextual ranking list (upper left) and aggregated SERP (upper right). Correlations are observed between the blue-links and modules marked by the same color, and the detailed informative terms are listed.}
\label{fig:CARL framework}
\end{figure}

Conventional aggregated search frameworks use a pipeline with two sequential subtasks: module selection and module presentation. 
The first subtask selects relevant modules from all feasible verticals,
and the second decides where to place each selected module in a given contextual ranking list.
Here, the contextual ranking list, later referred to as \textit{context}, is a ranking list of blue-links and is provided by a general web search engine. 
Machine learning methods have been applied to module selection\cite{arguello2009sources} and module presentation\cite{arguello2010vertical}\cite{konig2009click}. 
However, these methods develop policies for each subtask independently while the correlation between blue-links and modules is ignored.  
Taking Figure 1 for an example, informative terms from the blue-links in the context imply that the video module is probably more relevant to the query.
However, this important \textit{context information} is ignored in existing module selection methods and the framework may fail to retrieve the desired modules.
Thus, the overall performance of aggregated search framework is not optimal by separately optimizing each subtask. 

Adapting \emph{Markov decision process} (MDP) for aggregated search provides a unified framework that helps alleviate the limitations of existing pipelines.
Recently, MDP-based models have shown strong capacity in information retrieval tasks\cite{wei2017reinforcement}\cite{xia2017adapting}\cite{oosterhuis2018ranking}. 
In these models, MDP formalizes a ranking task as sequential decision making where each time step corresponds to a ranking position and each action corresponds to selecting one document.
The model performance is optimized by learning a ranking policy that makes decision at each time step by computing probabilities of the action selection based on the current state.
To enable the use of information in different forms that helps action selection (e.g., preceding documents selected\cite{xia2017adapting}), state representation is designed to incorporate these information.
Inspired by the success of MDP in these ranking tasks, we propose that aggregated search can also be conducted by learning a ranking policy that iteratively selects ranking items (i.e., blue-links or modules) to construct SERP.  
Under the framework of MDP, the ranking policy can employ context information and is learned to optimize the end quality of aggregated SERP without intermediate subtasks.
However, learning such ranking policies for aggregated search is challenging.
On one hand, different verticals may correlate to the context information in different ways, which is difficult to model~\cite{sun2017collaborative}.
On the other hand, there is no trivial solution to utilize context information for heterogeneous item ranking since the feature sets of ranking items vary significantly. 

To address these challenges, we propose CARL, an MDP-based model with \textit{\underline{c}ontext \underline{a}ttention} and \textit{\underline{r}epresentation \underline{l}earning}.
Our model captures correlations between the context and verticals by a novel pseudo module component that encodes context information by a \emph{recurrent neural network} (RNN) \cite{WANG2019}\cite{xu2018spatio} with vertical-specific attention mechanism.
The encoded context information is incorporated into a \textit{learnable} module embedding that is formulated as a combination of module content embedding and the pseudo module.
The representation learning, which includes learning for content embedding and pseudo module, is jointly optimized with ranking policy through agent-environment interactions formulated by MDP.
As shown in Figure 2, at each time step (i.e., ranking position), the agent receives a state from the environment. 
Our model applies a GRU (RNN with gated recurrent unit) to encode the state and derive both state representations and module embeddings.
Based on the encoded state, the action (i.e., item selection) is decided by the ranking policy, which computes the probability of selection for each blue-link and module.
After taking the action, an updated state and a reward are received from the environment. 
Moving to the next iteration, the above process is repeated until the constructed SERP achieves a given target length.
The optimization function of MDP is defined as the cumulative rewards, which guides the joint learning to maximize the quality of the constructed SERP.

To achieve a more effective joint learning in CARL, we further propose an optimization function with auxiliary supervision. 
Compared with other MDP-based ranking models, CARL jointly learns module embeddings and ranking policy instead of applying preliminary representation settings (e.g., output vector of pretrained doc2vec\cite{xia2017adapting}).
Thus, an optimization function using only the cumulative rewards may not be sufficient for effective joint learning. 
To this end, we design two self-supervised passes to provide auxiliary supervision signals. 
As shown in Figure 2, at each time step, the self-supervised passes compute loss based on representations of encoded states, the learned pseudo module, and the action taken. 
The optimization function with the self-supervision loss is applied to further boost the joint learning.


In summary, this paper makes the following contributions:
\begin{itemize}
  \item We propose a novel model named CARL for aggregated search, which can effectively model and utilize the context information by joint learning of ranking policy and context-aware module embeddings.
  \item We propose an optimization function with self supervision loss, which makes the joint learning more effective in CARL by providing auxiliary supervision.
  \item Experiments using two public datasets show that CARL outperforms several baselines and the learning process is more effective with the proposed optimization function.
\end{itemize}

\section{RELATED WORK}

\begin{figure*}[htbp]
\centering


\includegraphics[width=7.2in]{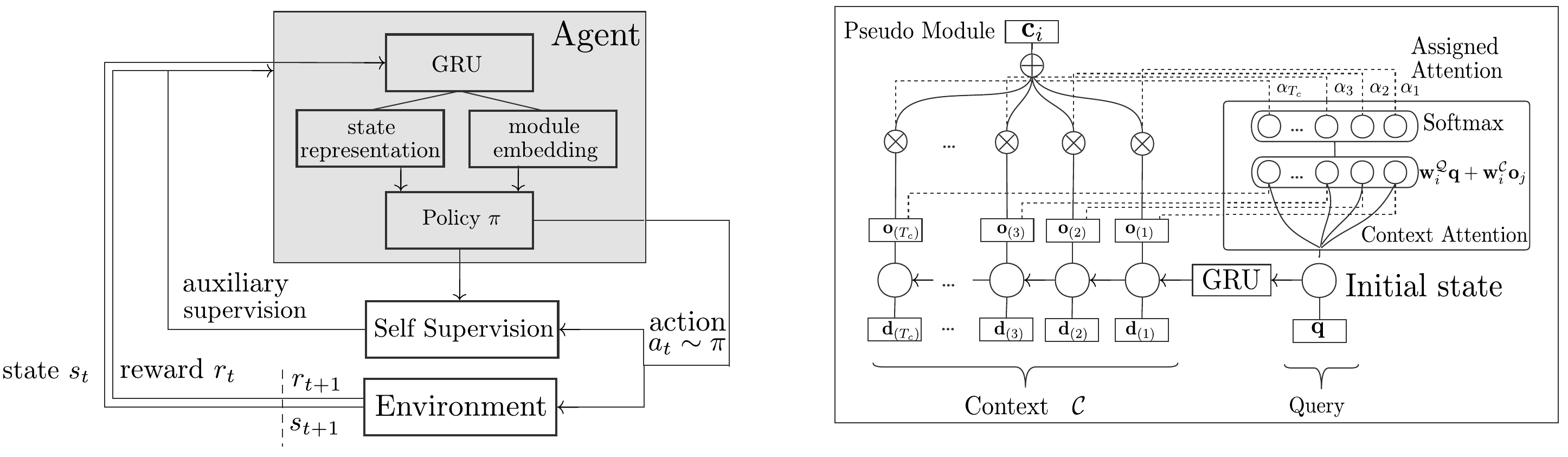}

\caption{ Overall architecture of CARL (left) and module embedding generation (right). Each time step the agent applies a GRU to encode the received state as state representation and module embeddings. An action is selected by the policy, and the agent is optimized by both rewards and auxiliary supervision.}

\label{fig:CARL framework}
\end{figure*}

\subsection{Aggregated Search}
Aggregated search refers to searching through a collection of specialized search engines, \textit{verticals}, and aggregating heterogeneous results (i.e., modules and blue-links) on the search result page (SERP). 
Many studies focus on developing approaches for the two subtasks in existing pipelines: module selection and module presentation\cite{arguello2017aggregated}.
The module selection task is usually regarded as a multiclass classification task, and methodologies developed range from heuristic single-evidence approaches \cite{arguello2009sources} to machine learning based approaches \cite{arguello2010vertical}\cite{konig2009click}. 
For example, a linear classifier is trained to predict the relevance of each vertical\cite{arguello2010vertical}, and a non-linear classifier GBDT is applied to conduct the prediction task\cite{konig2009click}.
The module presentation task is formulated as a re-ranking problem given the contents to present \cite{wang2016beyond}. 
Similar to learning-to-rank approaches, pointwise ranking functions \cite{arguello2011learning}\cite{jie2013unified} and pairwise preference estimators \cite{arguello2011learning}\cite{ponnuswami2011composition} are used to determine the final page layout.

The evaluation of aggregated SERP can be conducted by on-line methods \cite{wang2016beyond}\cite{jie2013unified} and test collection methods \cite{wang2016evaluating}\cite{wang2018joint}.
On-line methods are based on click through data from real users in a live environment. 
For example, a click-skip metric is applied to model user satisfaction\cite{wang2016beyond}.
Test collection methods use relevance label based metrics to evaluate SERP, where a wide range of metrics for ad-hoc retrieval tasks can be adopted (e.g, DCG~\cite{jarvelin2002cumulated} and ERR~\cite{chapelle2009expected}).  
Zhou~et al.\cite{zhou2012evaluating} proposes a utility-based framework to better approximate user satisfaction by extending traditional ad-hoc retrieval metrics. 



\subsection{Reinforcement Learning for Information Retrieval}

Reinforcement learning has been applied to solve many information retrieval (IR) tasks.
For example, by modeling document ranking as a sequential decision making problem, a ranking policy under Markov decision process (MDP) framework can be trained for search result diversification\cite{xia2017adapting} and relevance ranking\cite{wei2017reinforcement}.
A framework based on Deep Q-Network (DQN) is proposed to simultaneously optimize ranking content and display order for complex ranking settings when the optimal display order is not known a priori\cite{oosterhuis2018ranking}.
The multi-scenario ranking problem in E-commerce is modeled as a multi-agent sequential decision problem, and agents (i.e., ranking strategies in different scenarios) are collaboratively optimized to improve the overall performance\cite{feng2018learning}.
Reinforcement learning is also used in many recommendation tasks.
For example, an MDP-based collaborate filtering (CF) model is proposed to take into account the long-term effects of each recommendation.
DQN is leveraged to integrate users' negative feedback (e.g., skipped items) into the recommendation strategy and maximize the long-term cumulative rewards\cite{zhao2018recommendations}. 

\section{PRELIMINARIES}

\subsection{Problem Definition}
We consider the following setting: 
a query is issued to a general web engine and several vertical sub-engines; the returned results from the general web engine and the vertical sub-engines are lists of blue-links and modules, respectively. 

Blue-links are homogeneous documents that only contain textual information. 
Each query is associated with a set of blue-links:
$D = \{ \VECTOR d_1, \VECTOR d_2,...,\VECTOR d_K \}$.
Verticals are specialized search services (e.g., images, videos, Q\&A). The set of feasible verticals is predetermined and denoted as $ \mathcal{V} = \{v_1, v_2,..., v_J \}$. Each vertical $v_j $ is associated with a sub-engine that returns modules in response to a query. 
For ease of presentation, we denote the general web as $v_0$ to distinguish it from the verticals.

Each module is formed via aggregation of documents from the same vertical, and is presented as a ``block of documents" in the SERP.
Unlike blue-links, modules are heterogeneous because features available to different verticals vary significantly (e.g., thumbnail images for the image vertical and number of votes for the Q\&A vertical). 
We denote the complete set of module as 
$ M = \{\VECTOR m_1, \VECTOR m_2, ... ,\VECTOR m_J \} $, 
where $\VECTOR m_j $ is the module returned from vertical $v_j $.

Aggregated search is to construct a search result page (SERP) of target length from a candidate set 
$X=\{D,M\}$ including blue-links and modules. 
The item placed at each position is either a blue-link $\VECTOR d_i \in D $, or a module $\VECTOR m_j \in M $. 
The optimal \textit{policy} maximizes the utility of the SERP in terms of topical relevance and query intent.

\subsection{Markov Decision Process}
Markov decision process (MDP) provides a framework for modeling a sequential decision making process, where reinforcement learning can be applied to find the optimal policy from the agent-environment interaction.
An MDP with discrete action space and deterministic state transition is defined as quintuple $\langle \mathcal{S}, \mathcal{A}, \textit{T}, r, \pi \rangle$, and each component is:

\textbf{States:}  $\mathcal{S}$ represents a set of possible states that the process may be in at any time step. A state can contain multiple aspects of information that are useful for decision making.

\textbf{Actions:}  $\mathcal{A}$ is a discrete set of actions the agent can take. \mbox{The actions available at different time steps may be different. }

\textbf{Transition:}  $\textit{T}$ is the state transition function represented by $s_{t+1} = T(s_t, a_t) $, which determines the next state according to the action taken and the state at the previous time step.

\textbf{Reward:}  $r$ is the received utility of taking an action at a state and is represented as a reward function $r(s, a)$.

\textbf{Policy:} $\pi$ is the distribution over available actions based on the current state.
The probability of taking each action is determined by a policy function $ p(a|s)$ which maps the representation of states and actions to probabilities.

The agent and environment interact as follows: at each time step $t$, the agent receives a state $s_t \in \mathcal{S}$ from the environment, and according to policy $\pi(a|s_t)$, it selects an action $a_t \in \mathcal{A}$. 
The agent receives a numerical reward $r(a_t, s_t)$, and changes to the next state, $s_{t+1} = T(s_t, a_t)$. 
The agent aims to learn the optimal policy that maximizes the total received reward
\begin{displaymath}
  \sum_{t=0}^{T} \E_{\pi} [\gamma ^t r(s_t,a_t)]
\end{displaymath}
where $0 < \gamma \leqslant 1 $ is the discount rate, $T$ is number of steps during the interaction.
Thus, the decision making at each time step should account for both immediate and long-term rewards.

\section{THE PROPOSED CARL MODEL}

As discussed in the former section, it is challenging to model and utilize context information for aggregated search since interactions with context vary across verticals.
In this work, we propose a joint learning model named CARL that encodes the context information into module embedding learning and develop a ranking policy based on the learned embeddings.  
The overall structure of CARL is illustrated in Figure 2.

\subsection{Joint Learning under MDP Formulation}
The construction of the aggregated search result page can be considered as a process of sequential decision making with an MDP, in which each time-step corresponds to a ranking position, and an action corresponds to selecting an item to place at the current ranking position. 
The decision making relies on a policy function that estimates probabilities of an item being selected based on state and item representations.

Learning a unified policy function for heterogeneous items is challenging because of the \textit{incomparable} item representations.
Specifically, features available to different verticals may vary substantially and every feature is not equally predictive\cite{arguello2017aggregated}.
A preliminary representation represents heterogeneous items only with commonly available features~\cite{jie2013unified}.
We first adopt this setting to illustrate a basic unified solution under the MDP formulation. 
We will propose a model with embedding learning to more effectively address this challenge.

Based on module embedding learning, the candidate set of each query, which consists of all available ranking items, is represented as  $ X = \{ \VECTOR x_1, \VECTOR x_2 ,...,\VECTOR x_{K+J} | \VECTOR x_i \in \R^{\alpha} \} $, where $\{ \VECTOR x_i \}_{i=1,...,K}$ and $\{ \VECTOR x_i \}_{i=K+1,..., K+J} $ represents blue-links and modules, respectively. 
The states, actions, transition function, reward and policy in the MDP framework are formulated as follows:

\textbf{States:} State at time-step t is formulated as a quadruple $s_t = \langle \VECTOR q, \mathcal{Z}_t, X_t , \mathcal{C} \rangle $, where $\VECTOR{q}\in \R^{\alpha}$ is the representation of the issued query; $\mathcal{Z}_t = \{\VECTOR x_{(n) } \}_{n=1}^{t}$ is the partial ranking list so far, where $\VECTOR x_{(n)} $ is the representation of the item ranked at position $n$; 
$X_t $ is the set of candidate items;
$\mathcal{C} = \{\VECTOR x_{(n) } \}_{n=1}^{T_c}$ is contextual ranking list and $T_c$ is number of items in context.
The initial state is $s_0 = \langle\VECTOR q, \varnothing , X , \mathcal{C} \rangle $.   


\textbf{Actions:}  $\mathcal{A}(s_t) $ is the set of available actions at time step \textit{t}. The action $a_t \in \mathcal{A}(s_t)$ selects a blue-link or module $\VECTOR x_{m(a_t)} \in X_t $ for the ranking position $\textit{t}+1 $, where $m(a_t)$ is the index of the selected item. 
We also use $ v(a_t)$ to denote the corresponding vertical of the selected item $\VECTOR x_{m(a_t)}$.

\textbf{Transition:}  The state transition is deterministic, and the transition function \textit{T} is defined as:

\begin{displaymath}
\begin{aligned}
  s_{t+1} = 
  &\, T(s_t, a_t) = T([\VECTOR q, \mathcal{Z}_t, X_t,\mathcal{C}], a_t) \\
  =
  & [\VECTOR q, \mathcal{Z}_t \oplus \{\VECTOR x_{m(a_t)} \}, X_t \setminus \{\VECTOR x_{m(a_t)}\}, \mathcal{C}],
\end{aligned}
\end{displaymath}
where the selected item is appended ($\oplus$) to the partial ranking list $\mathcal{Z}_t$ and removed from the candidate set $X_t$. 
The process is terminated when the constructed ranking list $\mathcal{Z}_t $ reaches the target length or the candidate set $X_t $ becomes empty.

\textbf{Reward:}  We derive the reward function from a given evaluation measure for the constructed search result page. 
Different evaluation metrics to measure user satisfaction can be applied. 
We consider a cumulative gain-based metric nDCG\cite{jarvelin2002cumulated} and a utility-based general evaluation framework\cite{zhou2012evaluating} in this work. The details of these metrics will be presented in the experiment section.

Let the evaluation for the ranking list at time-step $t$ be $R(s_t)$. 
The reward for action $a_t$ is computed by:   
$r(s_t, a_t)=R(s_{t+1})-R(s_t)$, where $R(s_0)$ is 0.

\textbf{Policy} $\pi$: 
Since ranking is modelled as a sequential process, we use a recurrent neural network with gated recurrent unit (GRU) to encode the ranking so far and guide the item selection for the next position. 
The GRU maps a state $s_t = \langle\VECTOR q, \mathcal{Z}_t, X_t , \mathcal{C}\rangle$ to a real vector, where $ \mathcal{Z}_t = [\VECTOR x_{(1)},\VECTOR x_{(2)},...,\VECTOR x_{(t)} ]$ is the partial ranking list at time step $t$. 

The GRU is formulated as:
\begin{equation}
\begin{aligned}
\VECTOR u_t = 
& \sigma (\VECTOR W_u^x \cdot \VECTOR x_t + \VECTOR W_u^s \cdot \VECTOR o_{t-1}) \\
\VECTOR r_t = 
& \sigma(\VECTOR W_r^x \cdot \VECTOR x_t + \VECTOR W_r^s \cdot \VECTOR o_{t-1} ) \\
\VECTOR o_t = 
& (1-\VECTOR u_t)\odot \VECTOR o_{t-1}+\VECTOR u_t \odot \VECTOR s_t \\
\VECTOR s_t = 
& \text{tanh}(\VECTOR W^x \cdot \VECTOR x_t + \VECTOR W^s \cdot (\VECTOR r_t \odot \VECTOR o_{t-1})) \\
\end{aligned}
\end{equation}
where $\VECTOR o $ is initialized with the query $\VECTOR q $: 
\begin{displaymath}
  \VECTOR o_0 = \sigma (\VECTOR W^q \cdot \VECTOR q),
\end{displaymath}
$\odot $ is the element-wise product,
$\VECTOR u_t, \VECTOR r_t, \VECTOR o_t, \VECTOR s_t \in \R ^{\alpha} $ denote the update gate, reset gate, output vector and cell state vector, respectively, and $\VECTOR W_u^x, \VECTOR W_u^s, \VECTOR W_r^x, \VECTOR W_r^s, \VECTOR W^x, \VECTOR W^s, \VECTOR W^q \in \R^ {\alpha \times \alpha}$ are learnable parameters of the GRU.
The output vector $\VECTOR o_t$ and cell state vector $\VECTOR{s}_t$ at the \textit{t}-th cell are concatenated as the representation of encoded state:
\begin{equation}
  GRU(s_t) =\VECTOR h_t = [\VECTOR o_t, \VECTOR s_t]
\end{equation}
The policy function takes as input the encoded state and the items' representation, and outputs the probability for each feasible action $a \in \mathcal{A}(s_t)$ via a soft-max function:
\begin{equation}
      p(a|s_t) = \frac{\exp\{ \VECTOR x_{m(a)}^T \cdot \VECTOR U_p \cdot GRU(s_t)  \}}{ \sum_{a^\everymodeprime \in\mathcal{A}(s)} \exp\{ \VECTOR x_{m(a^\everymodeprime)}^T \cdot \VECTOR U_p \cdot GRU(s_t)  \} }
\end{equation}
where $\VECTOR U_p \in \R^{\alpha \times 2\alpha}$ is learnable parameters. The policy $\pi$ at state $s_t$ is the distribution over all feasible actions:
\begin{displaymath}
  \pi(s_t) = <p(a_1|s_t),...,p(a_{|\mathcal{A}(s_t)|  }|s_t)>
\end{displaymath}

The parameters are optimized for maximizing the total rewards:
\begin{equation}
    \max \E_{\pi}[\sum_{t=0}^{T}r_t]
\end{equation}
where $r_t$ is the reward received at time-step $t$. Here we choose the discount rate $\gamma =1$ so that the policy is trained to directly optimize the given evaluation metric.

\subsection{Context-Aware Module Embedding}

For `informative' module embedding learning, we formulate the embedding of a module as a combination of the content embedding and pseudo module. 
Content embedding preserves multiple fields of information, and is obtained by projecting the module's original feature set into a common feature space. 
Pseudo module captures the query's vertical intent by encoding the contextual ranking list via a vertical-specific context attention.

Next, we describe how content embedding and pseudo module are derived for context-aware module embedding.  
Suppose during the ranking process, module $\VECTOR m_i $ that contains $k$ documents from vertical $v_i $ is in the candidate set. 
By concatenating all the features from the $k$ documents, the original representation of this module is:
\begin{displaymath}
  \VECTOR m_i = [\VECTOR m_T ^{(1)}, \VECTOR m_V^{(1)},... ,\VECTOR m_T ^{(k)}, \VECTOR m_V^{(k)}, \VECTOR m_i^{S}]
\end{displaymath}
where $\VECTOR m_T ^{(j)} $ and $\VECTOR m_V ^{(j)} $ are the textual and non-textual information of the  \textit{j}-th document, and $\VECTOR m_i^{S} $ is the structural information of modules from vertical $v_i$. 
Let $\delta_i$ be the dimension of the original representation, which varies across verticals. 
The combination of the content embedding and pseudo document is then formulated as:
\begin{equation}
  \VECTOR x_i =[\VECTOR c_{i} + \VECTOR V_{i}\cdot \VECTOR m_i ]
\end{equation}
where $\VECTOR c_{i}\in \R^{\alpha} $ is the pseudo module for vertical $v_i $, and $\VECTOR V_{i} \in \R^{\alpha \times \delta_i} $ is a vertical-specific projection layer. 

To derive the pseudo document that captures context information of query intent, we encode the top ranking results and apply attention mechanism to the hidden states. 
Here, the top ranking results refer to the ranking list prior to the ranking process using the same policy function under the MDP framework, where the candidate set has only blue-links:
\begin{displaymath}
  \mathcal{C} = [\VECTOR d_{(1)}, \VECTOR d_{(2)},...,\VECTOR d_{(T_c)}].
\end{displaymath}
Here $\VECTOR d_{(i)} \in D $ is the blue-link selected at position $i$, and $T_c$ is the length of the contextual ranking list.  
To obtain a simpler model, the same GRU from policy learning is also applied to encode the contextual ranking list
\begin{equation}
  GRU(\mathcal{C})= [\VECTOR o_{(1)}, \VECTOR o_{(2)},...,\VECTOR o_{(T_c)}]
\end{equation}
where $\VECTOR o_{(j)} $ computed by Equation (1) is the output vector for the item ranked at the \textit{j}-th position.

Given the encoded context $GRU(\mathcal{C}) $, the pseudo document $\VECTOR c_i $ for vertical $v_i $ is defined as
\begin{equation}
  \VECTOR c_i = \boldsymbol{\alpha}_i \cdot GRU(\mathcal{C})= \sum_{j=1}^{T_c}\alpha_{ij} \cdot \VECTOR o_j 
\end{equation}
where $ \alpha_{ij}$ represents the attention weight of the pseudo document $\VECTOR c_i $ to document $\VECTOR o_j $ and is computed by
\begin{equation}
  \alpha_{ij} = \frac{\exp\{e_{ij} \}}{\sum_{k=1}^{T_c}\exp\{ e_{ik}\}}
\end{equation}
where
\begin{equation}
  e_{ij} =  \sigma(\VECTOR w_i^{\mathcal{Q}} \VECTOR q + \VECTOR w_i^{\mathcal{C}} \VECTOR o_j +b_i )
\end{equation}
and $\{ \VECTOR w_i^{\mathcal{Q}}, \VECTOR w_i^{\mathcal{C}}, b_i   \}_{i=1,..., J } $ is the parameters to blend the captured query intent, the context, and the module itself.

Considering both the content embedding and pseudo module, the overall policy function is:
for module
\begin{equation}
  p(a|s_t) = \text{softmax}\{ [\VECTOR c_{v(a)} +\VECTOR V_{v(a)}\cdot \VECTOR m_a ]^T \cdot \VECTOR U_p \cdot GRU(s_t)  \}
\end{equation}
for blue-link
\begin{equation}
    p(a|s_t) = \text{softmax}\{\VECTOR x^T_{m(a_t)} \cdot \VECTOR U_p \cdot GRU(s_t)  \}
\end{equation}

\subsection{Training with Auxiliary Supervision}

In our MDP formulation, the ranking policy is jointly optimized with the context-aware module embeddings learning.
It is difficult to guarantee the effectiveness of embedding learning in this end-to-end paradigm, where only the rewards received from the environment is used as supervision signals. 
In order to promote the learned module embeddings to be in the same semantic space as blue-links, we propose an optimization function with auxiliary supervision that are provided by two self-supervised models: inverse and forward passes.

The inverse pass takes as input the encoded partial ranking list (\emph{not} the contextual list) before and after an item is added, and predicts the source vertical of this newly added item. 
To well perform the prediction task, it requires not only the module embeddings well incorporate the vertical information, but in the form that could be decoded from the state representation computed by GRU. The inverse pass with parameters $\theta_I$ is defined as
\begin{equation}
 \hat v(a_t)  =\phi _{inv} (\VECTOR h_t, \VECTOR h_{t+1};\theta_I)
\end{equation}
where $\hat v(a_t)$ is the predicted distribution over all verticals $V$. The inverse pass is trained to minimize the cross-entropy loss
\begin{equation}
  \min_{\theta_I} L_I(v(a_t), \hat v(a_t)).
\end{equation}

In addition to the inverse pass, a forward pass is trained to regularize the embedding learning process. In the inverse pass, directly optimizing for the prediction loss may lead to a degenerate solution\cite{agrawal2016learning}, in which module embeddings of different verticals are in distinct distributions instead of a comparable semantic space. 
This degenerate solution is not wanted since a general ranking policy is applied. 
Thus, a forward model uses the embedding of module content and the pseudo module to predict the state representation at the next time step, i.e.,
\begin{equation}
\begin{aligned}
\hat h_{t+1, c} = 
& \phi _{fwd}(\VECTOR h_t, \VECTOR c_{v(a_t)})\\
\hat h_{t+1,v} = 
& \phi _{fwd}(\VECTOR h_t, \VECTOR V_{v(a_t)} \cdot \VECTOR m_a)\\
\end{aligned}    
\end{equation}
where $\VECTOR h_t$ is the encoded state, $\VECTOR c_{v(a_t)} $ is the pseudo module, $\VECTOR V_{v(a_t)} \cdot \VECTOR m $ is the embedding of module content. The forward pass with parameter $\theta_F $ is trained to minimize the L-2 norm of the two predicted hidden state,
\begin{equation}
 \min_{\theta_F} L_F(\hat h_{t+1, c}, \hat h_{t+1,v})
\end{equation}

Thus, the overall optimization function is:
\begin{equation}
  \max_{\theta_P, \theta_I, \theta_F} [\E_{\pi }[\sum_t r_t] - L_I- L_F]
\end{equation}
The policy gradient algorithm of REINFORCE\cite{sutton2018reinforcement} is adopted to jointly optimize the model parameters towards maximizing the optimization function.   


\section{EXPERIMENTS}

\subsection{Experiment Setting}

We conduct experiments using two public datasets: FedWeb13 and FedWeb14 derived from the TREC federated web search track\cite{demeester2014overview} in 2013 and 2014. 
Each year, the tasks take 50 official queries and collect the candidate documents for each query by sending the query to over 150 sub-engines of 24 verticals.
Table \ref{tab:verticals and engines} shows example verticals and sub-engines.
The tasks gather snippets and web-pages of up to 10 returned results from each sub-engine.
The relevance of documents for queries has five levels based on the information need.

We adopt the approach proposed by Bota et al.\cite{bota2014composite} to sample modules for each vertical based on the returned results.
We preprocess the textual information of blue-links and modules with porter stemming, tokenization and stop-words removal.
We further extract feasible non-text information for modules from corresponding web-pages, and adopt the same presentation features as used in \cite{wang2016beyond}.
For each experiment, we perform 5-fold cross validation.



We use two types of test collection evaluation metrics.
The first is nDCG\cite{jarvelin2002cumulated}, a widely used ad-hoc retrieval metric that is also used by the TREC FedWeb track result merging task.
The second is a utility-based evaluation framework for aggregated search result page\cite{zhou2012evaluating}. 
This method extends ad-hoc retrieval metrics and evaluates the qualities of SERPs based on both relevance and query orientation (i.e., query intent).
We use the AS$_{DCG}$ and AS$_{ERR}$ under this framework based on the browsing models DCG\cite{jarvelin2002cumulated} and ERR\cite{chapelle2009expected}, respectively.

\begin{table} [tbp]
\caption{ Examples of verticals and sub-engines in FedWeb13and14. 
} \label{tab:verticals and engines}
    \centering
\scriptsize


\begin{tabular}{lll}

\toprule
Vertical & Returned results & Sub-engines\\
\midrule
Image & online images   & Flickr, Photobucket\\
Video & online videos  & Youtube, AOL Video\\
News & news articles  & BBC, CNN\\
Q\&A & answers to questions   &Yahoo Answers, StackOverflow\\
... & ...  &... \\
Academic & research papers  & ArXir.org, CiteSeerX\\
Blog & blog articles  & Google Blogs, LinkedIn Blog\\
Encyclopedia & encyclopedic entries  &Wikipedia, \\
Shopping & product shopping page   &Amazon, eBay \\
\midrule
General web & blue-links  & -  \\
\bottomrule

\end{tabular}

\end{table}

\begin{table*} [tbp]
\caption{ Performance comparison of models trained on annotation labels (full supervision). 
} \label{full supervision}
    \centering

\scriptsize

\setlength\tabcolsep{4.0pt}

\begin{tabular}{l|llll|llll}
\toprule
& \multicolumn{4}{c|}{ Performance on FedWeb13 } & \multicolumn{4}{c}{ Performance on FedWeb14 } \\

Method & nDCG@5  & nDCG@10 & AS$_{DCG}@10$ & AS$_{ERR}@10$ & nDCG@5  & nDCG@10& AS$_{DCG}@10$ & AS$_{ERR}@10$  \\
\midrule
RankNet & 0.3641  & 0.3811  & 0.2972  &	0.2239 & 0.3811 &0.3913  &  0.3145&0.2433	  \\
LambdaMART & 0.3629   &	0.3794  &	0.2826 & 	0.2135 & 0.3742 &0.3932 &0.3296 & 0.2476\\
LTM & 0.3904   &	0.4153   &	0.3277  & 	0.2417 &0.4158 &0.4390 &0.3311 &0.2721 \\
\midrule
MDP-DIV(AS$_{\text{DCG}}$) & 0.4288  &	0.4375  &	0.3587  &	0.2439 &0.4572 &0.4376 &0.3763 &0.3135 \\
DRM(AS$_{\text{DCG}}$)& 0.4171   &	0.4292  & 	0.3648  &	0.2645 &0.4603&0.4801 &0.3786 &0.3200\\
\midrule
\TEXTBF{CARL(AS$_{\textbf{DCG}}$) }& 0.4520  &	0.4635$^{*}$  &	\TEXTBF{0.4327}$^{*}$  &	\TEXTBF{0.3327}$^{*}$ &0.5023$^{*}$ &0.5260$^{*}$ &\TEXTBF{0.4751}$^{*}$ & \TEXTBF{0.3818}$^{*}$\\
\TEXTBF{CARL(nDCG)}   & \TEXTBF{0.4644}$^{*}$ &	\TEXTBF{0.4824}$^{*}$ &	0.4269$^{*}$  &	0.3164$^{*}$  & \TEXTBF{0.5162}$^{*}$ & \TEXTBF{0.5398}$^{*}$ & 0.4629$^{*}$ & 0.3724$^{*}$\\
\bottomrule
\end{tabular}

\end{table*}

\begin{table*} [tbp]
\caption{ Performance comparison of models trained on generated click data (weak supervision). 
} \label{tab: weak supervison}
    \centering

\scriptsize

\setlength\tabcolsep{4.0pt}

\begin{tabular}{l|llll|llll}
\toprule
& \multicolumn{4}{c|}{ Performance on FedWeb13 } & \multicolumn{4}{c}{ Performance on FedWeb14 } \\

Method & nDCG@5  & nDCG@10 & AS$_{DCG}@10$ & AS$_{ERR}@10$   & nDCG@5  & nDCG@10 & AS$_{DCG}@10$ & AS$_{ERR}@10$ \\
\midrule
RankNet  &0.3149 &0.3247 &0.2617 &0.2144   &0.3414 &0.3744  & 0.2910&0.2541  \\
LambdaMART  &0.3234 &0.3310  &0.2642 &0.2096   &0.3522 &0.3713  &0.3082 &0.2562  \\
LTM &0.2747 &0.3156  &0.2583 & 0.1817 &0.3178 &0.3591 & 0.2877&0.2194 \\
\midrule
MDP-DIV  &0.2788 &0.2944  &0.2393 &0.1872  &0.2993 &0.3272  &0.2639 &0.2382 \\
DRM  & 0.2816& 0.3041  &0.2420 &0.2066  &0.3075 &0.3466  &0.2761 &0.2425 \\
\midrule
\TEXTBF{CARL (without SSL)} &0.3591$^{*}$ &0.3877$^{*}$  &0.2971$^{*}$ &0.2329$^{*}$   &0.3841$^{*}$ &0.4089$^{*}$  &0.3533$^{*}$ &0.2897$^{*}$ \\
\TEXTBF{CARL}    & \TEXTBF{0.4014}$^{*}$ &\TEXTBF{0.4243}$^{*}$  & \TEXTBF{0.3552}$^{*}$ &\TEXTBF{0.2784}$^{*}$  &\TEXTBF{0.4355}$^{*}$ &\TEXTBF{0.4461}$^{*}$  &\TEXTBF{0.4217}$^{*}$ &\TEXTBF{0.3316}$^{*}$ \\
\bottomrule
\end{tabular}

\end{table*}

We compare the CARL with the following baselines:

\textbf{RankNet}\cite{burges2005learning}: RankNet is a learning-to-rank (LTR) approach learns a scoring function using gradient decent with listwise losses.

\textbf{LambdaMART}\cite{wu2010adapting}: LambdaMART is also a LTR approach and is the state-of-the-art variant of RankNet.

\textbf{LTM}\cite{lee2015optimization}: Learning to Merge extends a data fusion technique, LambdaMerge, by modelling the impact of document quality and vertical quality separately.


\textbf{MDP-DIV}\cite{xia2017adapting}: MDP-DIV  develops the ranking policy by modelling the dynamic utility user perceives when browsing the ranking list.

\textbf{DRM}\cite{oosterhuis2018ranking}: DRM  uses Deep Q-Network to estimate the state value after adding a document and conducts a greedy policy to maximize the total rewards of the ranking list.



Note that MDP-DIV was originally introduced for search result diversification, and DRM for complex ranking setting. 
By changing the reward functions, these methods can be optimized to also solve the aggregated search task.
This setting is also used by Oosterhuis et al.\cite{oosterhuis2018ranking} in the experiments.

We consider two settings of training supervision for model comparison.
The first is \textit{full supervision} using target evaluation metrics (e.g., nDCG) as supervision based on annotation labels. 
The second is \textit{weak supervision} in the form of click-through data. 
Training under this setting is more closed to real-world scenarios where click-through data are more easily obtained compared to fully annotated document set.
We adopt the procedure proposed by Joachims et al. \cite{joachims2017unbiased} to generate click data for FedWeb13 and FedWeb14.
This procedure includes two phases. 
First, we train all baselines using full supervision and select the one with best performance (MDP-DIV for FedWeb13 and DRM for FedWeb14) as a base ranker. Second, we use the base ranker to generate click-through data by simulating browsing process of users: each item is clicked only if it is observed and perceived as relevant.


The baselines using the MDP framework (i.e., MDP-DIV, DRM) need preliminary representations of queries and candidates for training. 
We use the output representation of a pretrained doc2vec\cite{le2014distributed} to encode the textual information of queries and candidate items. 
The setting and dataset we use for pretraining the doc2vec model follow the same procedure as Xia et al.\cite{xia2017adapting} for direct comparison.

\subsection{Results of Performance Comparison}

We conduct experiments using two training settings, i.e., full supervision and weak supervision, on datasets FedWeb13 and FedWeb14.
Boldface indicates the highest score among all approaches. 
We denote significant improvements of CARL over the best baseline results with `$^{*}$' in tables (based on two-tailed paired t-test and p-value $<$ 0.05).
The overall performance comparison of the baselines and our model are shown in Table \ref{full supervision} and \ref{tab: weak supervison}. 

For training with full supervision, we consider two variants of our proposed model, denoted as \TEXTBF{CARL(AS$_{\textbf{DCG}}$) } and \TEXTBF{CARL(nDCG)}, where rewards are AS$_{DCG}$ and nDCG, respectively.
We can see from Table \ref{full supervision} that CARL consistently outperforms all baselines in all evaluation metrics by a substantial margin.
Compared with DRM, which is the most competitive MDP-based baseline, CARL achieves more than 28.1\%, 38.2\% improvements on nDCG@5, nDCG@10 and 45.8\%, 32.0\% improvements on AS$_{DCG}$@10, AS$_{ERR}$@10.
For two different variants of our model, they achieve the best performance on the corresponding evaluation metrics (nDCG for CARL(nDCG), AS$_{\text{DCG}}$ and AS$_{\text{ERR}}$ for CARL(AS$_{\text{DCG}}$)).
This result indicates that CARL can optimize the quality of SERP in terms of different \mbox{metrics by using the metric as rewards.}

For training with weak supervision, we can see from Table \ref{tab: weak supervison} that CARL still maintains the superiority over baselines although sparse supervisions lead to performance decreases of all methods.
We also include the results of CARL trained without self-supervision loss (SSL), in which the self-supervised models (Section 4.3) are not used and the optimization function only consists of the received feedbacks.
Compared with CARL trained without SSL, CARL achieves more than 11.2\%, 13.0\% improvements on nDCG@5, nDCG@10 and 20.3\%, 18.9\% improvements on AS$_{DCG}$@10, AS$_{ERR}$@10.
This performance boost indicates that our optimization function with self-supervision loss can provide essential auxiliary supervision for a effective joint learning when the feedbacks received from outside are not sufficient enough. 

\subsection{Effect of CARL Components}
\begin{table} [tbp]
\caption{ Effects of context sampling policy. 
} \label{tab:context sampling}
    \centering
\scriptsize

\setlength\tabcolsep{3.0pt}

\begin{tabular}{l|l|lll}

\toprule
& Quality of  & \multicolumn{3}{c}{ Performance of  } \\
& context & \multicolumn{3}{c}{ aggregated search } \\
\midrule
Context & nDCG@10 & AS$_{DCG}@5$ & AS$_{DCG}@10$ & nDCG@10  \\
\midrule
No context & - & 0.3464 & 0.3519 & 0.4211 \\
Random context & 0.1263 & 	0.3521 & 	0.3642 & 0.4324\\
LTM &0.3532 &0.3774 &0.3943 &0.4621 \\
RankNet & 0.3715 &	0.3849 &	0.4098 & 0.4709\\
LambdMART  & 0.3857 &0.4182 &0.4187 & 0.4783\\
CARL & 0.4221 &	\TEXTBF{0.4378}$^{*}$ & \TEXTBF{0.4463}$^{*}$	&\TEXTBF{0.4984}$^{*}$  \\

\bottomrule
\end{tabular}

\end{table}

\begin{table} [tbp]
\caption{ Effects of GRU choice. 
} \label{tab:GRU choice}
    \centering
\scriptsize

 \setlength\tabcolsep{4.0pt}

\begin{tabular}{lll}

\toprule
GRU choice & AS$_{DCG}@5$ & AS$_{DCG}@10$  \\
\midrule
dual-GRU (without SSL) & 	0.3851 & 	0.3967\\
dual-GRU  &	0.4109 &0.4327	\\
\midrule
uni-GRU (without SSL)  & 0.3695 & 0.3886 \\
uni-GRU  &	\TEXTBF{0.4178}$^{*}$ &	\TEXTBF{0.4609}$^{*}$ \\
\bottomrule
\end{tabular}
\end{table}

\subsubsection{Effect of Context}
We first investigate the effects of different context sampling methods to the performance of our model. 
In CARL, the context (i.e., contextual ranking list) is sampled from blue-link set by the same ranking policy to aggregated SERP.
Besides reusing the same policy, we can apply different ranking methods for context sampling. 

Table \ref{tab:context sampling} presents the test performance of different ranking policies for retrieving contextual ranking lists.
We can see from the experiment result that a high-quality context can provide beneficial query intent information for the aggregated search policy.
Without the contextual ranking list (\textit{No context}), or with low-relevance context (\textit{Random context}), the performance of the CARL framework is not as good as the result with a high-relevance context.
We can also see that although the context by LTM or LTR approaches is of relative low relevance (measured by nDCG@10), our model can still achieve comparable quality of the generated SERP.
Besides contextual ranking list, various types of information (e.g., users activities\cite{sun2016contextual}\cite{sun2016collaborative}) can be considered into context in the future.

\subsubsection{Effect of GRU Choice}
We study the benefit of using only one GRU for underlying representations sharing.
The GRU is applied in two places: encoding the blue-links in contextual ranking list (c.f. Eq (6)); and encoding the variable partial ranking list during aggregated SERP generation process (c.f. Eq (2)). 
We can instantiate the same GRU in these two places, while another option is to use two unique GRUs.

Table \ref{tab:GRU choice} shows the experiment results of using single or dual GRUs. 
To investigate the effects of optimization function with self-supervision loss (SSL), we also compare results without SSL.
We can see from Table \ref{tab:GRU choice} that when the model parameters are trained without SSL, dual-GRU outperforms single-GRU possibly because one single-GRU is difficult to encode the representations of the contextual ranking list and partial ranking list in the same semantic space.
Single-GRU achieves the best performance, and this supports the argument that the learned context-aware embeddings are \textit{comparable}, since we can effectively apply the same GRU to encode both the contextual list and the heterogeneous partial ranking list.

\section{CONCLUSION}
We proposed a model named CARL for aggregated search based on Markov decision process (MDP).
In contrast to modularized solutions that perform aggregate search through subtasks, CARL directly optimizes the quality of the constructed SERPs by a joint learning of the ranking policy and context-aware module embeddings.
The learned embeddings for heterogeneous modules are optimized to be both informative and comparable using the optimization function with auxiliary supervision. 
Experiments using public TREC datasets showed that CARL significantly outperforms the baseline approaches in multiple evaluation metrics. 


\section*{Acknowledgement}
We would like to thank Xiaojie Wang for his help.
This work is supported by China Scholarship Council (CSC) Grant \#201808240008,
Australian Research Council (ARC) Discovery Project DP180102050,
Overseas Cooperation Research Fund of Graduate School at Shenzhen, Tsinghua University (Grant No. HW2018002).

{
\bibliographystyle{IEEEtran}
{
\bibliography{ijcnn.bib}
}
}

\end{document}